\documentclass{osa-article}

\journal{osajournal}

\articletype{Research Article}

\usepackage{lineno}

\begin{document}

\title{Computational 3D microscopy with optical coherence refraction tomography}

\author{Kevin C. Zhou,\authormark{1,}* Ryan P. McNabb,\authormark{2} Ruobing Qian,\authormark{1} Simone Degan,\authormark{3} Al-Hafeez Dhalla,\authormark{1} Sina Farsiu,\authormark{1,2} and Joseph A. Izatt\authormark{1,2,+}}

\address{\authormark{1}Department of Biomedical Engineering, Duke University, Durham, NC 27708, USA\\
\authormark{2}Department of Ophthalmology, Duke University Medical Center, Durham, NC 27710, USA\\
\authormark{3}Department of Radiology, Duke University Medical Center, Durham, NC 27710, USA}

\email{\authormark{*}kevin.zhou@duke.edu} 
\email{\authormark{+}jizatt@duke.edu}



\begin{abstract*}
Optical coherence tomography (OCT) has seen widespread success as an in vivo clinical diagnostic 3D imaging modality, impacting areas including ophthalmology, cardiology, and gastroenterology. Despite its many advantages, such as high sensitivity, speed, and depth penetration, OCT suffers from several shortcomings that ultimately limit its utility as a 3D microscopy tool, such as its pervasive coherent speckle noise and poor lateral resolution required to maintain millimeter-scale imaging depths. Here, we present 3D optical coherence refraction tomography (OCRT), a computational extension of OCT which synthesizes an incoherent contrast mechanism by combining multiple OCT volumes, acquired across two rotation axes, to form a resolution-enhanced, speckle-reduced, refraction-corrected 3D reconstruction. Our label-free computational 3D microscope features a novel optical design incorporating a parabolic mirror to enable the capture of 5D plenoptic datasets, consisting of millimetric 3D fields of view over up to $\pm75^\circ$ without moving the sample. We demonstrate that 3D OCRT reveals 3D features unobserved by conventional OCT in fruit fly, zebrafish, and mouse samples.
\end{abstract*}



\newcounter{video}
\newcommand{\genvidlabel}[1]{%
  \refstepcounter{video}\label{#1}%
}

\genvidlabel{video:zebrafish}
\genvidlabel{video:fruitfly}
\genvidlabel{video:esophagus}
\genvidlabel{video:trachea}

\section{Introduction}
First introduced 30 years ago, optical coherence tomography (OCT) \cite{huang1991optical} has since evolved into a broad class of 3D imaging techniques based on low-coherence interferometry that has impacted a variety of fields, including ophthalmology, cardiology, and gastroenterology. OCT owes much of its success to its coherent detection mechanism, attaining near shot-noise-limited imaging performance and enabling high-rate 3D volumetric imaging with millimeter-scale depth penetration in scattering tissues without optical clearing \cite{wieser2014high, carrasco2017review}.

However, this same detection strategy is also the source of OCT’s most notable limitations – poor lateral resolution due to its tradeoff with the depth of focus (DOF), and coherent speckle noise that can be similar in magnitude to the desired signal \cite{karamata2005speckle}, arising in part from the band-pass transfer function of OCT in 3D $k$-space \cite{zhou2021unified}. Existing DOF-extension approaches, such as beam-shaping \cite{ding2002high, leitgeb2006extended, lee2008bessel, liu2011imaging}, suffer from loss in signal-to-noise ratio (SNR) due to backcoupling inefficiencies. Further, digital refocusing techniques, such as interferometric synthetic aperture microscopy (ISAM) \cite{ralston2007interferometric}, also lose SNR away from the nominal focus and, as coherent synthesis techniques, require phase-stable measurements. On the other hand, previous angular compounding speckle reduction approaches \cite{desjardins2006speckle} have incorporated only limited angular ranges, thus restricting their effectiveness. Furthermore, wavefront-modulation approaches \cite{liba2017speckle} can degrade resolution and SNR. These longstanding limitations of OCT degrade the interpretability and effectiveness of its contrast mechanism, compared to incoherent microscopy techniques, and ultimately limit the diagnostic utility of OCT. 

Here, we present 3D optical coherence refraction tomography (OCRT), a new computational volumetric microscopy technique that extends OCT, featuring a multi-angle incoherent $k$-space synthetic reconstruction algorithm. 3D OCRT thus not only exhibits the coherent detection sensitivity advantages of OCT, but also exhibits a speckle-free incoherent contrast mechanism analogous to that of incoherent microscopy, together with multifold enhanced lateral resolution over an extended 3D field of view (FOV). The key innovations of 3D OCRT are two-fold. First, we experimentally demonstrate a novel optomechanical design featuring a parabolic mirror as the imaging objective, with which we were able to acquire OCT volumes from multiple views over up to $\pm75^\circ$ without moving the sample. More generally, our approach is the first experimental demonstration of a more general class of conic mirror-based methods that can in principle acquire images from multiple views over up to $\pm90^\circ$ across two rotation axes using low-inertia scanners (e.g., galvanometers) as the only moving parts \cite{zhou2021high}. Existing approaches have achieved much smaller angular ranges \cite{carrasco2015pupil} or required mechanically rotating the imaging optics \cite{yao2015angular}. Our work is thus a generalization of our previous work on 2D OCRT, which demonstrated substantial improvements over conventional OCT in 2D through single-axis sample rotation \cite{zhou2019optical}. Here, we demonstrate the capture of 5D plenoptic datasets (3D space + 2D angle) without moving the sample itself, generating a wealth of data from which new sources of 3D contrast can be computationally synthesized. 

To handle these large 5D datasets, our second key innovation is a novel computationally efficient 3D reconstruction algorithm that leverages differential programming frameworks and optimization techniques developed in the deep learning community \cite{abadi2016tensorflow}. In particular, our approach allows dense 3D reconstruction from simultaneous participation of all multi-angle OCT volumes across arbitrarily large 5D datasets (in our case, $\sim$90 GB), using a single memory-limited graphics processing unit (GPU). This algorithm is a substantial improvement over our 2D OCRT reconstruction algorithm \cite{zhou2019optical}, whose large memory requirement precluded GPU use, even on datasets that were several orders of magnitude smaller ($\sim$100 MB).

\section{3D OCRT}

\begin{figure*}[ht]
    \centering
    \centerline{\includegraphics[width=1.3\textwidth]{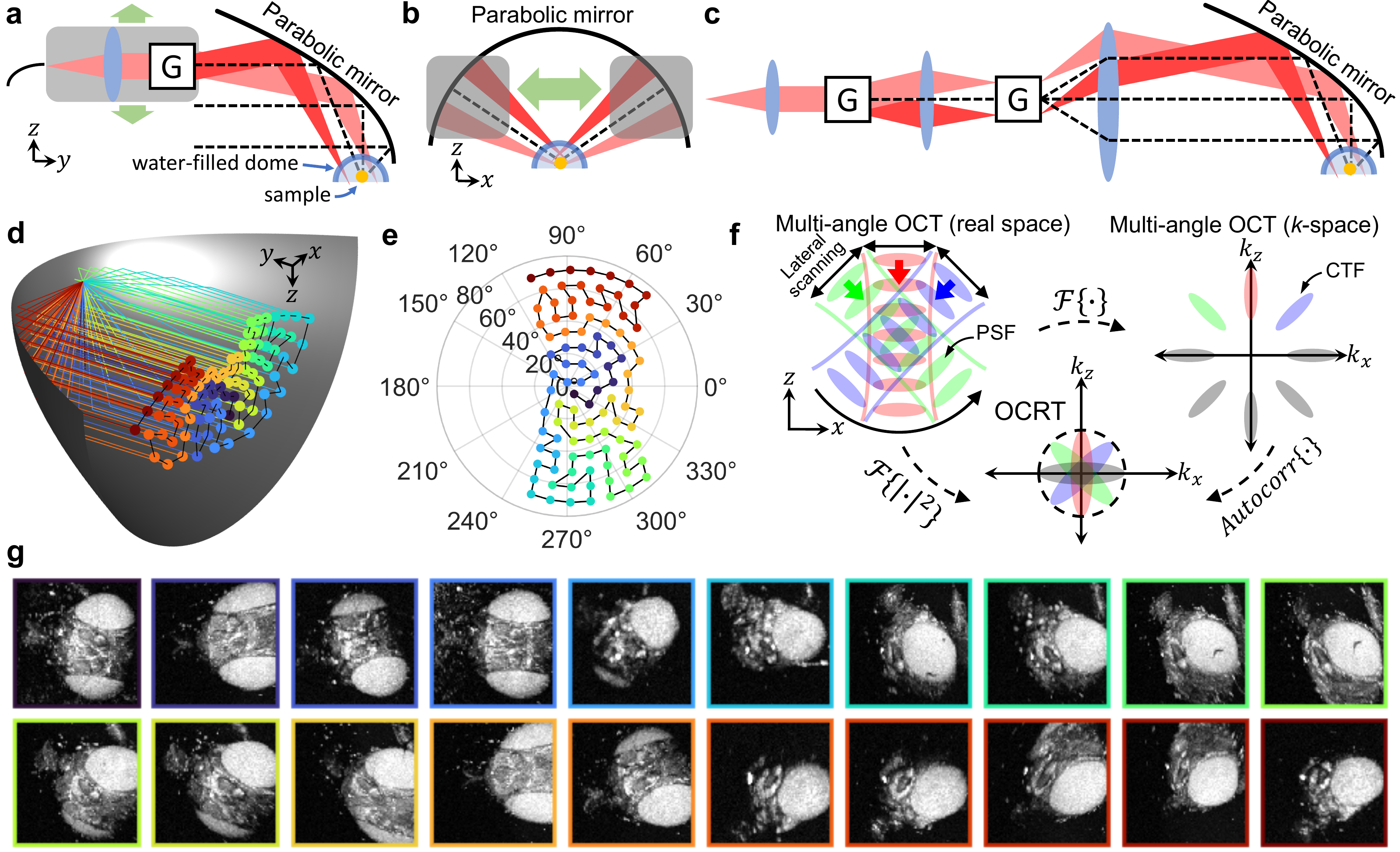}}
    \caption{Novel parabolic-mirror-based imaging system enables multi-view imaging over wide angular ranges for OCRT. (a-b) Schematic of our parabolic-mirror-based OCRT imaging system, featuring a 2D translating probe that angularly scans a collimated beam with galvanometers (G). The sample is placed at the center of a water-filled glass optical dome. (c) The same result can be achieved without 2D translation using another set of galvanometers. (d) 3D visualization of the angular coverage by a parabolic mirror via central chief rays for each of the 96 lateral probe positions that we used in our experiments. Rays are color-coded by order of acquisition, from purple to red. (e) The corresponding azimuthal and inclination angles of each central chief ray in (d). (f) \textit{k}-space theory of OCRT. (g) Example en face projections of a 5D OCRT dataset of a fruit fly, color-coded by incidence angle, consistent with (d) and (e). The lateral FOV of each projection is $\sim$1.7 mm\textsuperscript{2}.
    }
    \label{fig:theory}
\end{figure*}

\subsection{Incoherent 3D $k$-space synthesis with OCRT}
Although OCT is a coherent imaging modality characterized by a band-pass coherent transfer function (CTF) \cite{zhou2021unified}, OCRT differs from 2D \cite{alexandrov2006synthetic,zheng2013wide} and 3D \cite{lauer2002new} coherent synthetic aperture techniques in that the multi-angle OCT volumes are combined \textit{incoherently}, that is, by discarding the phase and only operating on intensity (Fig. \ref{fig:theory}f). As a result, the band-pass CTF effectively becomes demodulated down to a low-pass \cite{zhou2021unified}, which can be understood via the Wiener-Khinchin theorem, by which the magnitude-squaring of the OCT image in real-space corresponds to autocorrelation in $k$-space. As a result, the complex exponentials windowed by the band-pass OCT CTF are rephased to a common origin (i.e., DC), so that OCRT doesn't have the phase stability requirement that many other \textit{coherent} synthetic aperture approaches have.

Since these demodulated CTFs, or incoherent transfer functions (ITF), overlap at the $k$-space origin, they can be combined to form an expanded ITF when the OCT resolution is anisotropic (Fig. \ref{fig:theory}f). This is often the case, as the lateral resolution is typically $>$10 \textmu m, while the axial resolution can essentially be tuned independently via the source properties and can be submicrometer \cite{povazay2002submicrometer}. Thus, the lateral resolution increases monotonically with angular coverage (Fig. S1). In the limit of full angular coverage ($\geq$180$^\circ$), the synthesized ITF is isotropic and given by the original OCT axial resolution (or lateral resolution, whichever is better \cite{zhou2020spectroscopic}). Finally, because the observed speckle pattern decorrelates as a function of angular separation, we observe significant speckle reduction because of the incoherent angular compounding.

\subsection{Plenoptic imaging with parabolic mirrors}
To obtain this resolution enhancement, we require OCT volumes acquired over a very wide angular range, ideally without requiring sample rotation to maximize the generality of OCRT. To this end, we replaced the more typical refractive convex imaging objective with a reflective concave parabolic mirror, which allows independent 4D control of sample-incident the 2D lateral position and 2D angle (azimuthal and inclination) via scanners placed conjugate and anticonjugate to the sample \cite{carrasco2015pupil} (Fig. \ref{fig:theory}a-c). Parabolic mirrors are infrequently used for imaging because of their tilt-induced aberrations that restrict FOV, often necessitating sample translation \cite{lieb2001high, ruckstuhl2004attoliter}. However, we have exploited the quadratic dependence of FOV on lateral spot size when imaging in an off-axis configuration \cite{zhou2021high} to obtain millimetric FOVs with a lateral resolution of $\sim$15 \textmu m, consistent with conventional OCT systems. We note that this quadratic scaling with lateral resolution is identical to that of the DOF, meaning that tilt aberrations of parabolic mirrors do \textit{not} add additional FOV constraints on top of the lateral-resolution-DOF tradeoff we seek to circumvent \cite{zhou2021high}. Our experimental setup also features a water-filled optical dome placed at the mirror's focus, where the sample is positioned (Fig. \ref{fig:theory}a,b), to substantially reduce spherical aberrations that would otherwise occur when a focused beam refracts obliquely across a flat RI discontinuity (e.g., coverslip) interface \cite{zhou2021high}.

To test the feasibility of this novel use of parabolic mirrors and optical domes, we translated an angularly scanning collimated beam across the mirror's half aperture to vary the sample-incidence angle (Fig. \ref{fig:theory}a,b), noting that the same effect can be achieved rapidly with another pair of galvanometers (Fig. \ref{fig:theory}c). With this setup, we were able to obtain plenoptic measurements of the samples, resulting in 5D datasets consisting of approximately $1.3\times1.3\times1.65$ mm\textsuperscript{3} volumes over $\pm75^\circ$ and $\pm25^\circ$ about the $y$- and $x$-axes, respectively (Fig. \ref{fig:theory}d,e). Example en face OCT images from a 5D dataset of a fruit fly, projected across the $z$ dimension, are shown in Fig. \ref{fig:theory}g.

\subsection{Large-scale, joint OCT volume registration and computational 3D reconstruction}
Given this 5D plenoptic dataset, the goal of 3D OCRT is to register and superimpose the multi-angle volumes to realize the incoherent 3D $k$-space synthesis and speckle reduction described earlier. The registration algorithm jointly optimizes two sets of parameters: 1) sample-extrinsic, or those describing the positions and orientations of the sample-incident rays, as governed by the imaging system, and 2) sample-intrinsic, or those describing deformation of ray trajectories due to the spatially varying RI. 

In theory, the sample-extrinsic parameters are determined by the parabolic mirror's sole parameter (i.e, its focal length), the entry positions across the mirror aperture, and the angular scan amplitude (for lateral scanning across the sample), and can thus be modeled through analytical ray tracing. In practice, we account for imperfections or misalignments by allowing a separate set of optimizable parameters for each multi-angle OCT volume, controlling the sample-incident angle, telecentricity, lateral scan range, and field curvature, among others. The sample-extrinsic parameters generate the boundary conditions for ray propagation through the sample's 3D RI distribution, which is coaligned with the 3D backscatter-based reconstruction. For more details on the registration model, see extended methods in Supplementary Note 1. 

Both the sample-intrinsic and sample-extrinsic parameters are optimized via stochastic gradient descent in TensorFlow \cite{abadi2016tensorflow} by minimizing the error between the A-scans of each volume and their forward predictions, as well as regularization terms operating on the 3D RI distribution to promote smoothness and enforce object support. This inverse optimization algorithm of OCRT differs from that of many other inverse problems in that the resolution-enhanced, speckle-reduced reconstruction is not itself a directly optimizable parameter, but rather it is generated through superposition of all OCT volumes, akin to the backprojection algorithm of X-ray computed tomography (CT). However, since requiring joint participation of the entire 5D dataset at every gradient descent iteration, though possible in our original 2D implementation \cite{zhou2019optical}, would be computationally infeasible, we propose a stratified batching approach that incrementally accumulates the A-scan contributions to the reconstruction voxels jointly with the registration. Thus, for example, in some cases, the parameters may already be fully optimized and OCT volumes registered before the reconstruction is completely formed. For a more detailed description, see extended methods in Supplementary Note 1.
\begin{figure*}[t]
    \centering
    \centerline{\includegraphics[width=1.4\textwidth]{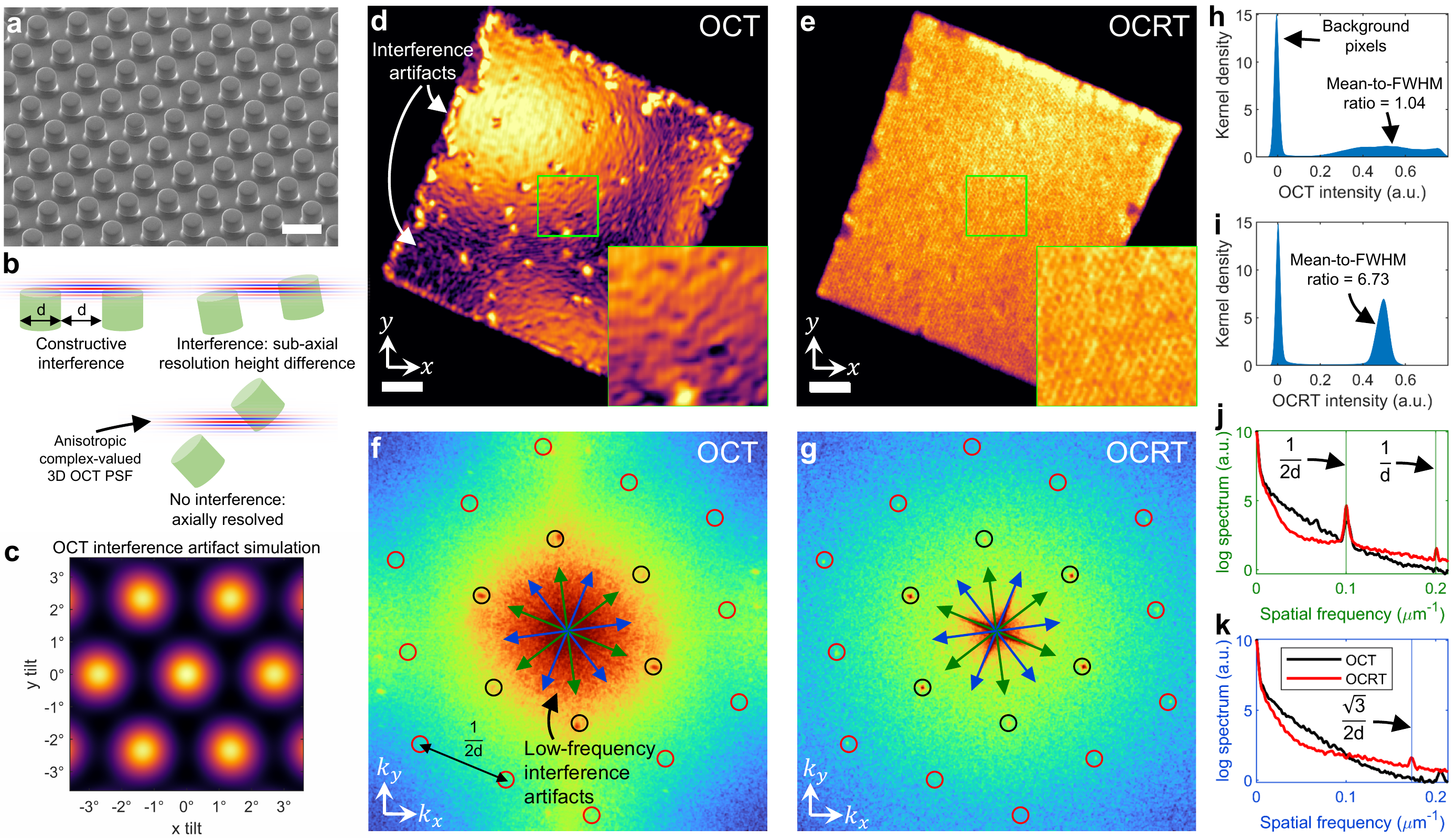}}
    \caption{OCRT enhances resolution and reduces interference artifacts (e.g., speckle) of OCT. (a) Scanning electron microscopy of a d = 5-\textmu m, hexagonally-arranged microstamp sample \cite{microstamps}. (b) Sub-resolution differences in pillar heights lead to interference. (c) En face OCT simulation of hexagonal lattice under various tilts predicts interference artifacts. See Fig. S2 for additional simulations. (d) En face OCT image (bilinearly upsampled to match OCRT) does not resolve the pillars and exhibits interference artifacts. (e) 3D OCRT reduces the artifacts and better resolves the pillars. (f) Log power spectral density (PSD) of OCT (black circles: fundamental frequency, $1/2d$; red circles: second harmonics, $1/d$ and $\sqrt{3}/2d$). (g) Log PSD of OCRT. (h) Kernel density estimate (KDE) of (d) exhibits a broad distribution of intensity values due to interference artifacts. (i) KDE of (e), however, exhibits a tight distribution due to speckle reduction. (j) Averaged 1D cross sections of (f) and (g) along green arrows. (k) Averaged 1D cross sections of (f) and (g) along blue arrows. Scale bars, 10 \textmu m in (a), 100 \textmu m in (d), (e).}
    \label{fig:microstamps}
\end{figure*}

\section{Results}

\subsection{Validation of lateral resolution enhancement and speckle reduction}
We first validated the resolution-enhancing and speckle-reducing capabilities of 3D OCRT by imaging a polydimethylsiloxane (PDMS) microstamp sample, consisting of hexagonally-arranged, 5-\textmu m-diameter, 5-\textmu m-tall cylindrical micropillars with an edge-to-edge spacing of 5 \textmu m (Fig. \ref{fig:microstamps}a). Since our OCT system had an axial resolution of 2.1 \textmu m and lateral resolutions of 15.3 ($x$) and 14.6 ($y$) \textmu m (or an anisotropy of $\sim$0.14) from a single view, it does not laterally resolve the 5-\textmu m pillars (Fig. \ref{fig:microstamps}d). Furthermore, this OCT image exhibits interference artifacts due to the fact that multiple pillars are probed by the PSF volume (Fig. \ref{fig:microstamps}b), consistent with simulated OCT responses to a hexagonal array as a function of tilt (Fig. \ref{fig:microstamps}c). Specifically, depending on the local sample tilt or non-telecentricity of the lateral scanning, the resultant axial separation of the pillars can lead to constructive or destructive interference (Fig. \ref{fig:microstamps}c), a direct consequence of the axial modulation in the 3D OCT PSF \cite{zhou2021unified}. See also Fig. S2 for OCT predictions matching experimental data, based on fitting-based estimates of microstamp surface normals. This is the same mechanism that underlies speckle formation, which is the interference result of a large number of sub-resolution scatterers. 

The 3D OCRT reconstruction much better resolves the pillars and eliminates the interference artifacts of OCT (Fig. \ref{fig:microstamps}e). The lateral resolution improvement over OCT can be further appreciated in the power spectra (Fig. \ref{fig:microstamps}f,g), in which more of the expected hexagonally-spaced peaks appear in OCRT than OCT, especially the second-harmonic peaks corresponding to the 5-\textmu m features (red circles in Fig. \ref{fig:microstamps}f,g). Fig. \ref{fig:microstamps}j,k show averaged 1D cross-sections of Fig. \ref{fig:microstamps}f,g along the blue and green arrows, with the expected Fourier peaks indicated with vertical lines. From these 1D plots, it is clear that OCRT contains the expected second-harmonic peaks, while OCT does not. The reduction of interference artifacts is also apparent, as Fig. \ref{fig:microstamps}f exhibits strong low-frequency artifacts that are absent in Fig. \ref{fig:microstamps}g.
Interference artifact reduction by OCRT is further quantified in Fig. \ref{fig:microstamps}h, i, which show the distribution of intensity values of Fig. \ref{fig:microstamps}d,e, where that of OCRT is $\sim$6.5$\times$ narrower than that of OCT.

The resolution enhancement results are consistent with theoretical predictions based on Fig. S1. In particular, we expected synthesized $x$ and $y$ lateral resolutions of $\sim$2.4 \textmu m and $\sim$6.6 \textmu m, corresponding to $\pm75^\circ$ and $\pm25^\circ$, respectively, indicating that our 3D OCRT reconstruction should resolve the 5-\textmu m pillars in the $x$ dimension, but not in the $y$ dimension. Indeed, in Fig. \ref{fig:microstamps}g, the red-circled peaks (corresponding to 5-\textmu m features) closer to the $k_x$-axis are stronger than those closer to the $k_y$-axis.

\begin{figure*}[!ht]
    \centering
    \centerline{\includegraphics[width=\textwidth]{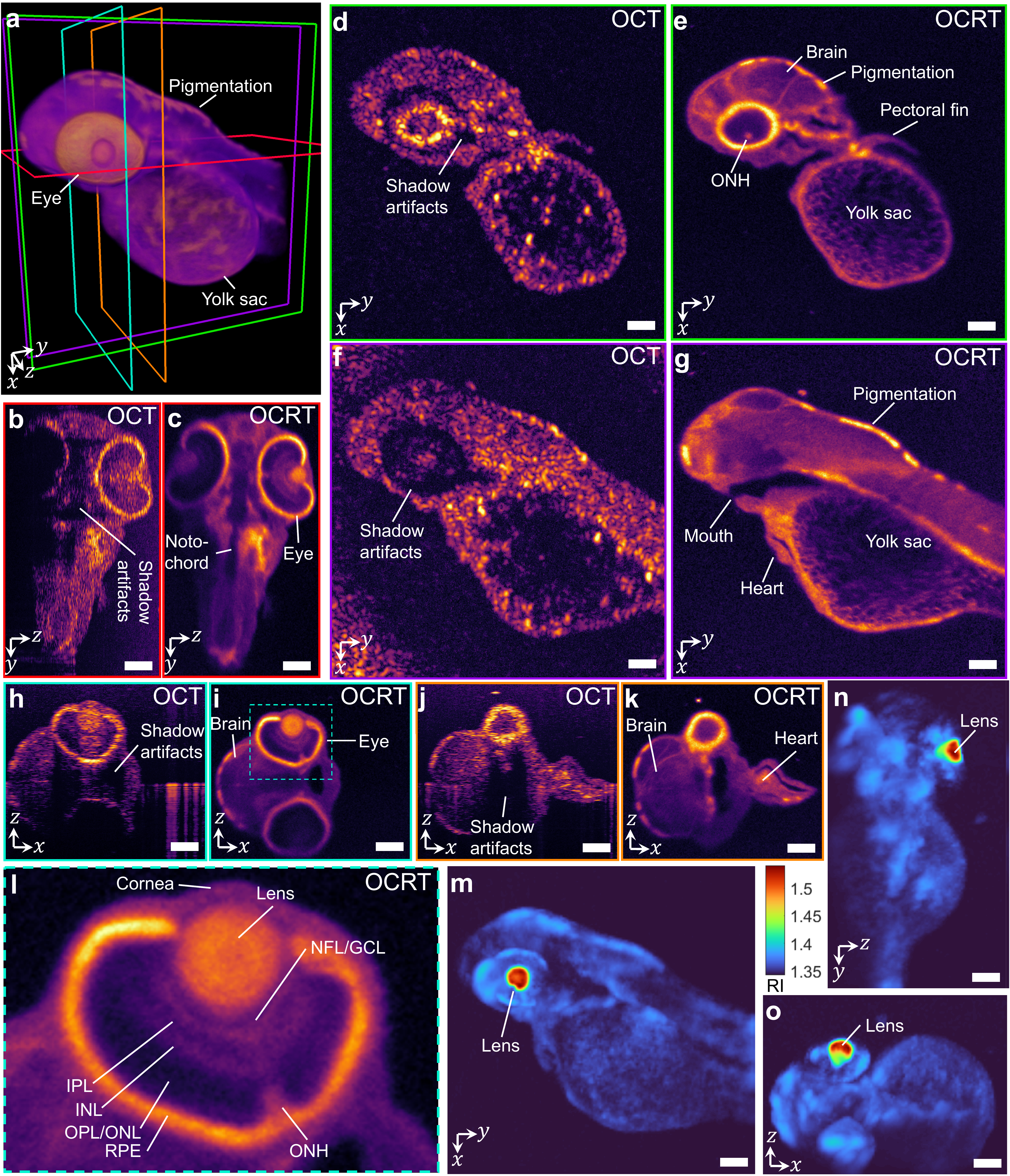}}
    \caption{Comparison of conventional OCT and 3D OCRT reconstruction of a 2-dpf zebrafish larva. (a) 3D rendering of OCRT, with color-coded slice locations of cross-sections in subsequent panels. (b-c) Comparison of a $yz$ slice. (d-g) Comparison of $xy$ slices at two different depths. (h-k) Comparison of $xz$ slices at two different $y$ positions. (l) Zoom-in of the eye in (i). NFL: nerve fiber layer, GCL: ganglion cell layer, IPL: inner plexiform layer, INL: inner nuclear layer, OPL: outer plexiform layer, ONL: outer nuclear layer, RPE: retinal pigment epithelium, ONH: optic nerve head. All OCT slices are histogram-matched to the corresponding OCRT slices. (m-o) Maximum intensity projections of the OCRT RI map. Scale bars, 100 \textmu m. See Visualization \protect\ref{video:zebrafish} for a full 3D comparison.}
    \label{fig:2dpf}
\end{figure*}

\begin{figure*}[ht]
    \centering
    \centerline{\includegraphics[width=1.1\textwidth]{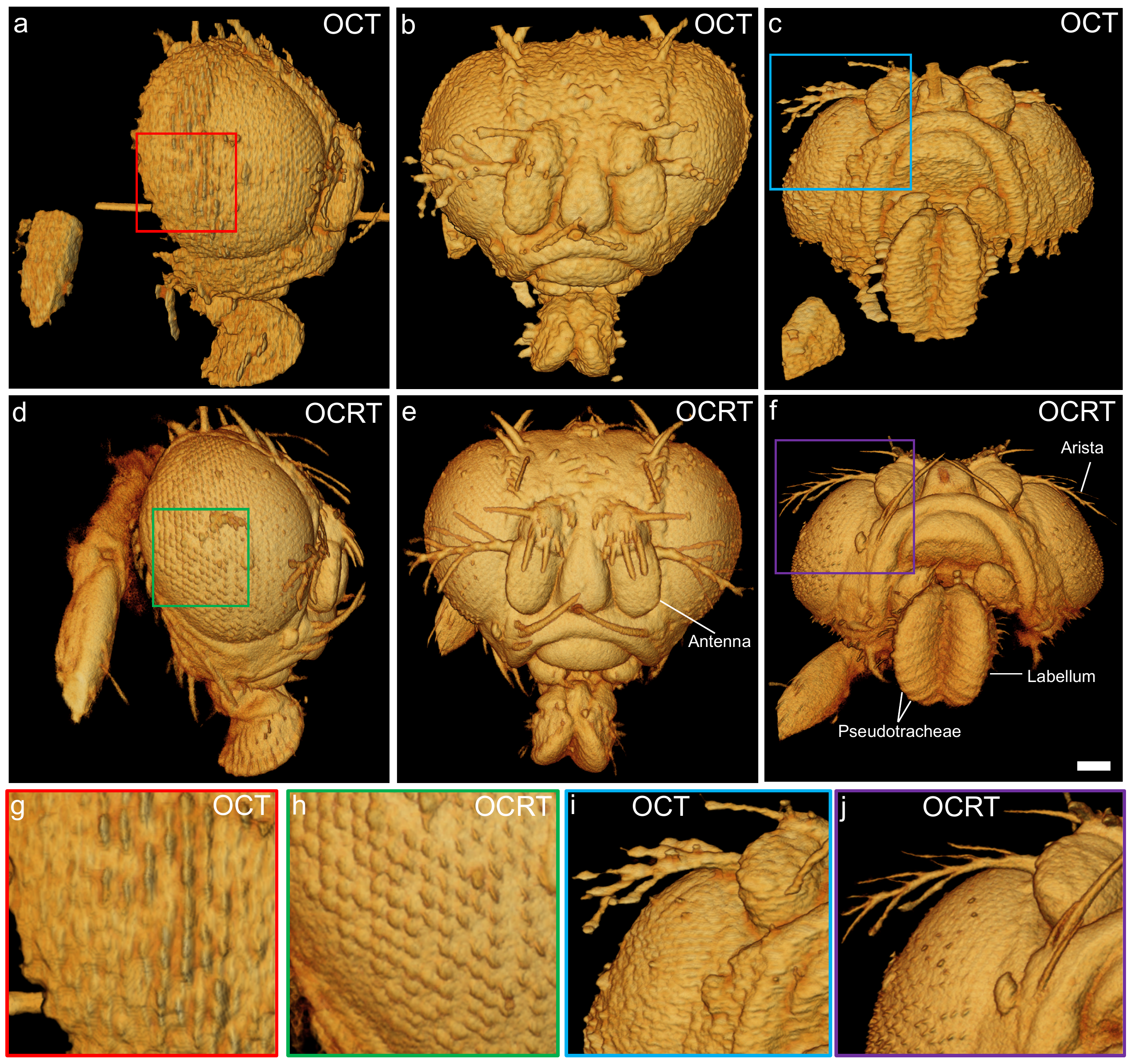}}
    \caption{Comparison of conventional OCT (a-c) and 3D OCRT reconstruction (d-f) of a fruit fly head. (a,d) Side view, with zoom-ins (g) and (h), highlighting the clearer reconstruction of the hexagonally-packed lenslets (ommatidia). (b,e) Front view ($z$-axis pointing towards the reader), highlighting clearer bristle reconstructions. (c,f) Bottom view of the 3D renderings, which shows clear reconstruction of the ridges (pseudotracheae) of the labellum and arista (i,j). Scale bar, 100 \textmu m. See Visualization \protect\ref{video:fruitfly} for a full 3D comparison.}
    \label{fig:fly}
\end{figure*}

\begin{figure}
    \centering
    \centerline{\includegraphics[width=1.3\textwidth]{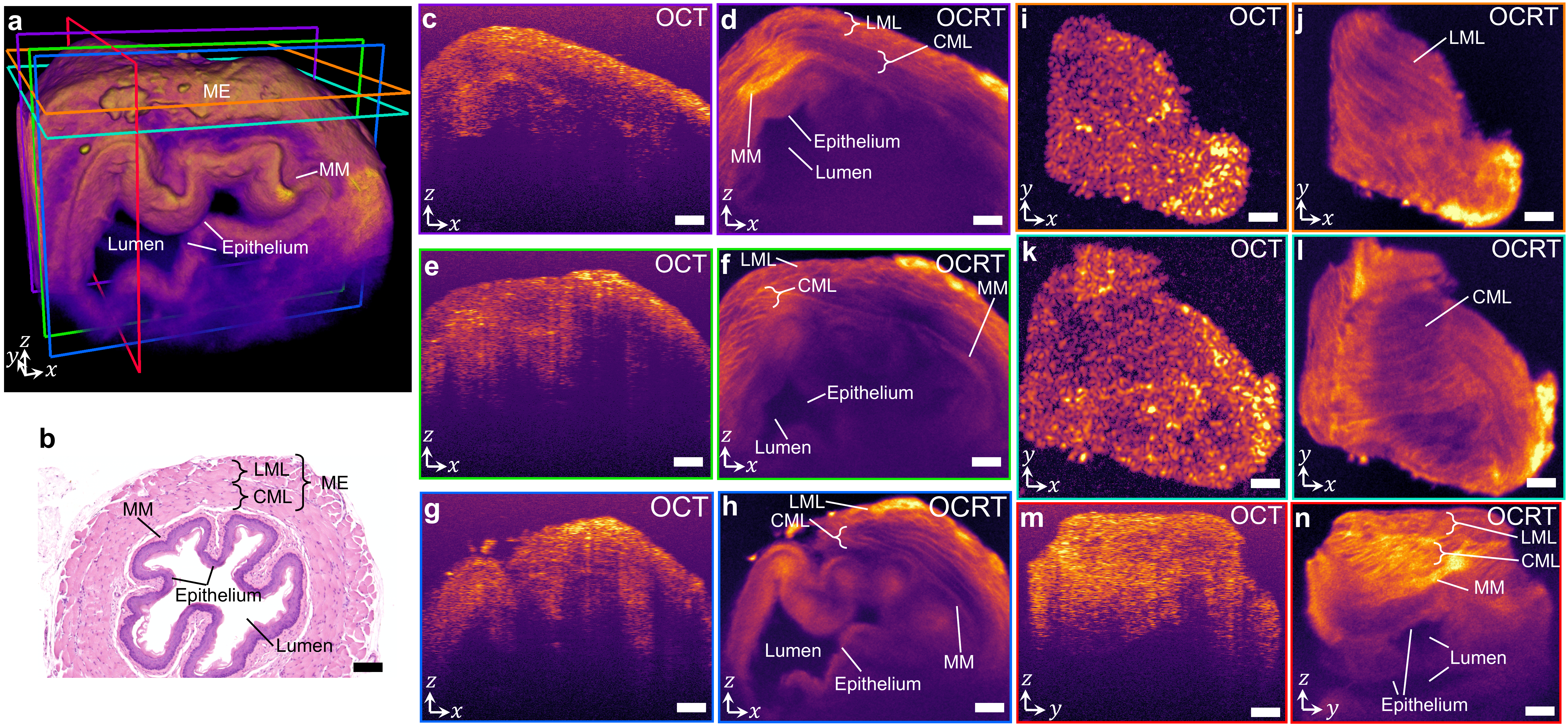}}
    \caption{Comparison of OCT and 3D OCRT of mouse esophagus. (a) 3D OCRT rendering, with color-coded slice locations of cross-sections in subsequent panels (MM: muscularis mucosae). (b) H\&E-stained histological section. The muscularis externa consists of a longitudinal muscle layer (LML) and a circular muscle layer (CML). (c-h) Comparison of various $xz$ slices. (i-l) Comparison of $xy$ slices st two depths, corresponding to the LML and CML. (m-n) Comparison of $yz$ slices. All OCT slices are histogram-matched to the corresponding OCRT slices. Scale bars, 100 \textmu m. See Visualization \protect\ref{video:esophagus} for a full 3D comparison.}
    \label{fig:esophagus}
\end{figure}

\begin{figure}
    \centering
    \centerline{\includegraphics[width=1.3\textwidth]{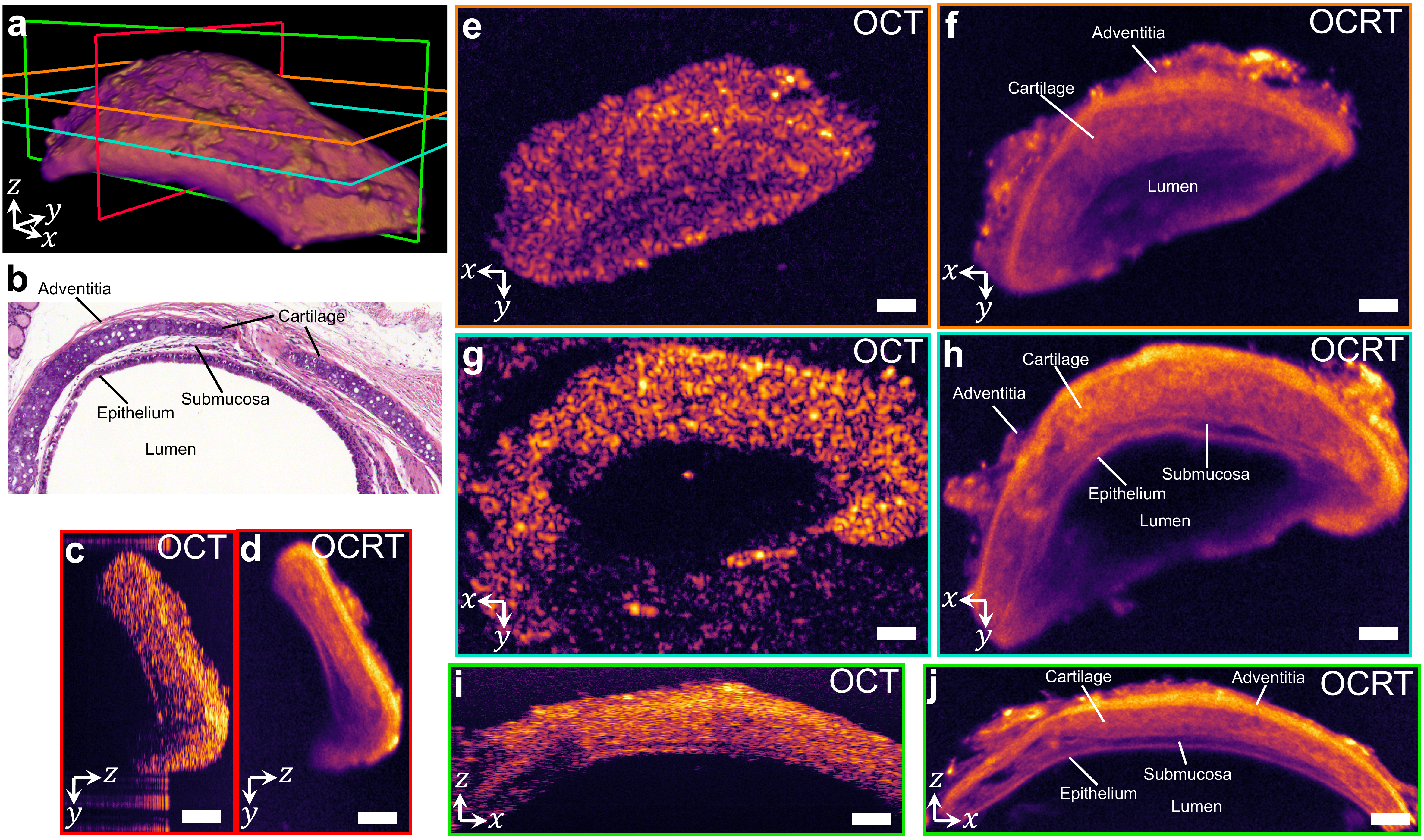}}
    \caption{Comparison of OCT and 3D OCRT of mouse trachea. (a) 3D rendering of OCRT, with color-coded slice locations of cross-sections in subsequent panels. (b) H\&E-stained histological section. The subsequent panels compare $yz$ slices (c,d), $xy$ slices (e-h), and $xz$ slices (i,j). All OCT slices are histogram-matched to the corresponding OCRT slices. Scale bars, 100 \textmu m. See Visualization \protect\ref{video:trachea} for a full 3D comparison.}
    \label{fig:trachea}
\end{figure}

\subsection{Biological results}
To demonstrate the generality of our new 3D OCRT implementation, we imaged and reconstructed several fixed samples: a zebrafish larva at 2 days post fertilization (dpf) (Fig. \ref{fig:2dpf}, Visualization \ref{video:zebrafish}), the head of an adult fruit fly (Fig. \ref{fig:fly}, Visualization \ref{video:fruitfly}), and various mouse tissue (Figs. \ref{fig:esophagus}, \ref{fig:trachea}; Visualizations \ref{video:esophagus}, \ref{video:trachea}). All samples were embedded in 2\% agarose (w/v) and immersed in water. 

In all cases, the 3D OCRT reconstructions offered substantial improvements over conventional OCT, owing not only to both the lateral resolution enhancement and speckle reduction, but also enhanced penetration depth, despite our imaging system not having access to both sides of the sample (in contrast to our original demonstration \cite{zhou2019optical}). Even in relatively transparent samples like zebrafish larvae, the speckle noise in OCT obscures many features that are revealed in OCRT (Fig. \ref{fig:2dpf}, Visualization \ref{video:zebrafish}). The improvements are especially apparent in the en face slices through the head and yolk sac of the zebrafish larvae, whose original OCT resolution is poor in both dimensions (Fig. \ref{fig:2dpf}d-g). Fine reticular structures in the yolk sac unresolvable by OCT are apparent in 3D OCRT. OCT also exhibits strong shadowing from the eye, as most directly apparent in Fig. \ref{fig:2dpf}b,h,j. This results in artifacts such as a dark ring around and below the base of the 2-dpf zebrafish larva's eye (Fig. \ref{fig:2dpf}d,f) that is recovered by OCRT (Fig. \ref{fig:2dpf}e,g). OCRT also reveals retinal layers and the optic nerve head (Fig. \ref{fig:2dpf}e,i,l), which are not apparent in OCT (Fig. \ref{fig:2dpf}d,h). Visualization \ref{video:zebrafish} shows a full flythrough comparison of the 3D OCRT and OCT volumes. Finally, the reconstructed RI maps of OCRT indicate a highly refractive (n $>$ 1.5) lens (Figs. \ref{fig:2dpf}m-o), consistent with previous findings \cite{greiling2008transparent}.

OCRT applied to an adult fruit fly (Fig. \ref{fig:fly}, Fig. S3, Visualization \ref{video:fruitfly}) also shows substantial improvement over OCT. Thanks to the speckle reduction and lateral resolution enhancement, the hexagonal packing of the individual micro lenslets (ommatidia) of the compound eye (see Fig. S3 and Supplementary Note 2) and the bristles (hairs) and aristae (branched bristles extending from the antennae) are better resolved by 3D OCRT. OCT, however, exhibits artificially bright and dark signals in the bristles and ommatidia, which are coherent interference artifacts not present in OCRT. OCRT also resolves the pseudotracheae on the labellum (tip of the extension from the mouth) (Fig. \ref{fig:fly}f). 

3D OCRT also offers significant improvements over OCT in mouse tissue, such as esophagus (Fig. \ref{fig:esophagus}, Visualization \ref{video:esophagus}). Notably, OCRT reveals the muscle fibers of the muscularis externa (ME), which consists of two layers -- the outer longitudinal muscle layer (LML) and the inner circular muscle layers (CML), which can be distinguished in the en face depth slices in Figs.  \ref{fig:esophagus}j and  \ref{fig:esophagus}l by the change in muscle fiber orientations. These two layers are also visible in the $xz$ cross-sections (Fig.  \ref{fig:esophagus}d,f,h). This enhanced visualization is attributable to both speckle reduction and resolution enhancement, as obliquely-oriented fibers cannot be resolved by poor OCT lateral resolutions. Below the ME, we can identify the muscularis mucosae (MM), epithelium, and lumen, especially in the cross-sectional cuts of the esophagus shown in Fig. \ref{fig:esophagus}d,f,h,n, consistent with hematoxylin and eosin (H\&E)-stained histological sections (Fig.  \ref{fig:esophagus}b). The MM is the thin hyperreflective layer in between the CML and the epithelium. All of these layers are very difficult to identify in the OCT images (Fig.  \ref{fig:esophagus}c,e,g,i,k,m). The improvement of OCRT over OCT is especially obvious in the flythroughs in Visualization \ref{video:esophagus}, in which the difference in muscle fiber orientations of the LML and CML is very clear for OCRT.

3D OCRT also offers substantial improvements over conventional OCT on mouse trachea (Fig. \ref{fig:trachea}), revealing several layers not readily apparent in OCT, most notably the hyaline cartilage rings, featuring lacunae or small cavities. We can also identify the outer adventitial layer (hyperreflective) as well as the submucosal (hyporeflective) and epithelial layers. The large speckle grains in OCT obscure these layers. Visualization \ref{video:trachea} shows a full flythrough comparison between 3D OCRT and OCT.

Finally, while all the 3D OCRT reconstructions presented so far were formed by taking the mean backscattered signal across all multi-angle views, other operations on the 5D OCRT datasets can yield new label-free information about the sample, which we discuss in Supplementary Note 3. For example, computing the variance across the angular dimensions yields a 3D OCRT reconstruction with orientational contrast \cite{lujan2011revealing}, highlighting structures within the yolk sac of the zebrafish, muscle fibers in the mouse esophagus, and cartilage in the mouse trachea (Fig. S4).

\section{Discussion}

We have presented 3D OCRT, a new computational volumetric imaging technique that substantially improves the image quality of OCT volumes through lateral resolution enhancement and speckle reduction. Furthermore, we have demonstrated a novel use of parabolic mirrors for multi-view imaging over very wide angular ranges without rotating the sample. Although parabolic mirrors are well known to exhibit ``perfect'' focusing only when the incident beam is parallel to the mirror's optic axis due to tilt aberrations, and therefore rarely used as imaging objectives, we demonstrated millimetric FOVs using the weakly-focused, long-DOF beams preferred in OCT. Since the 3D FOV generated by multi-view imaging would be limited by the DOF anyway, the limited lateral FOV of parabolic mirrors, having the same quadratic scaling with lateral resolution as the DOF, do not further restrict the 3D FOV. Thus, OCRT has a resolution advantage compared to optical projection tomography (OPT) for the same 3D FOV, because OCRT decouples resolution from the DOF or FOV. 

These improvements over conventional OCT make OCRT competitive with other incoherent (e.g., fluorescence-based) 3D microscopy approaches, such as multiphoton \cite{zipfel2003nonlinear} and light-sheet \cite{ahrens2013whole,chen2014lattice} microscopy. OCRT also inherits the many advantages of OCT, such as the longer near-infrared wavelengths typically used, which have higher penetration depths into scattering tissue. Further, while other point-scanning techniques rely on the narrow DOF of high-NA objectives for optical sectioning (e.g., confocal gating), OCT uses coherence gating, which has been shown to more strongly reject out-of-focus and light and therefore have better optical sectioning capabilities \cite{izatt1994optical}. Thus, high-NA objectives are not necessary for high-resolution 3D imaging with OCRT, potentially allowing for longer working distances and less sensitivity to aberrations. Specifically, even though in theory similar rays are used by both high-NA microscopy and OCRT, the former requires all multi-angle rays to be present at the same time to constructively interfere to form a focus. Any rays distorted in amplitude or phase (e.g., by occlusions and aberrations) would thwart the formation of such a focus. However, OCRT uses multi-angle rays sequentially, relying far less on their interference. Thus, OCRT's imaging depth is less affected by occlusions and aberrations, as evidenced in the zebrafish reconstructions below the highly scattering eye (Fig. \ref{fig:2dpf}).

Finally, because we are analyzing OCT and OCRT incoherently using ray-based models, we draw connections to concepts developed in the computer vision community, thus potentially opening new lines of investigations. For example, the 5D OCRT dataset has some similarities to 5D plenoptic function\cite{adelson1991plenoptic} from the field of light field imaging, which the describes the radiance as a function of two angular dimensions across 3D space. One difference is that the plenoptic function is often used to describe imaging of passively illuminated objects, as in photography, whereas OCT actively illuminates the object and observes the 180$^\circ$-backscattered light. As such, the 5D OCRT dataset also bears resemblance to the 6D spatially-varying bidirectional reflectance distribution function (SV-BRDF)\cite{nicodemus1965directional}, which measures radiance as a function of input and output illumination angles (2D each) across an opaque 2D manifold surface. OCRT, however, measures a degenerate version version of the SV-BRDF for the case of equal input and output angles, thus losing two dimensions, while gaining another dimension by measuring this information over 3D instead of 2D space. The two output angle dimensions can be obtained by modifying the OCT system to angle-resolve the back-scattered light, as is done in angle-resolved low-coherence interferometry\cite{wax2002cellular}. Thus, a method based on our parabolic mirror imaging system or other conic-section mirror-based imaging system\cite{zhou2021high} could lead to faster methods to acquire plenoptic light field or SV-BRDF data for other computational imaging applications.  

In summary, 3D OCRT is a new label-free, computational microscopy technique that yields a resolution-enhanced, speckle-reduced reconstruction and a coaligned 3D RI map that reveal new information not apparent in conventional OCT in a wide variety of biological samples. With conceptually straightforward improvements, in particular using faster sources, replacing 2D translation with anti-conjugate galvanometers, and deriving new forms of image contrast from the multi-angle data, 3D OCRT could see wide use in vivo biomedical imaging for basic scientific and diagnostic applications.

\begin{backmatter}
\bmsection{Funding}
This work was funded by the National Science Foundation (CBET-1902904) and the National Institutes of Health (U01EY028079, P30EY005722).

\bmsection{Acknowledgments}
We thank Z. Kupchinsky and N. Katsanis for providing the fixed zebrafish samples and Ramona Optics for providing the fruit fly sample. We also thank F. Wei for assisting with hardware construction.

\bmsection{Disclosures}
KCZ, RPM, RQ, AHD, SF, and JAI are inventors on patent applications related to the technologies described in this report.

\bmsection{Data Availability Statement}
Data underlying the results presented in this paper are available in TBD.

\end{backmatter}

\bibliography{sample}

\end{document}


\maketitle

\newcounter{video}
\newcommand{\genvidlabel}[1]{%
  \refstepcounter{video}\label{#1}%
}

\genvidlabel{video:zebrafish}
\genvidlabel{video:fruitfly}
\genvidlabel{video:esophagus}
\genvidlabel{video:trachea}

\section{Extended methods}
\subsection{3D OCRT imaging}
The imaging system is based on a commercial spectral domain OCT system (Bioptigen Envisu R4110 XHR SDOIS), whose illumination source is centered at 820 nm. Each single-view OCT volume consisted of 400$\times$400$\times$2048 unaveraged voxels, covering a 3D FOV of approximately 1.3$\times$1.3$\times$1.65 mm\textsuperscript{3} in water, acquired at a 20-kHz A-scan rate. The OCT sample arm optics were replaced with a tube lens pair (a pair of achromatic doublets, $f$ = 30 mm, 100 mm, Thorlabs), a protected-aluminum-coated parabolic mirror ($f$ = 12.5 mm, Edmund Optics), and a fused silica optical dome (outer radius = 8.128 mm, inner radius = 6.858 mm, CLZ Optics). The parabolic mirror was cut in half to facilitate sample mounting and inverted so that the cut surface was facing up. The optical dome was also inverted, centered at the parabolic mirror's focus, and filled with water as the immersion medium. The power incident at the sample was $\sim$1.5 mW, after $\sim$13\% loss upon reflection from the parabolic mirror. Resolution (FWHM) was characterized using sub-resolution polystyrene beads embedded in 2\% agarose (w/v); the axial resolution was 2.1 \textmu m, while the resolution in the lateral dimensions was 15.3 and 14.6 \textmu m ($x$ and $y$), enabling millimetric depths of field and lateral FOVs.

To vary the sample-incident angle, the sample arm components from the fiber output up to and including the tube lens (the probe) were mounted on two orthogonal translation stages (Zaber Technologies, Sigma Koki), which varied the 2D entry position across the aperture of the parabolic mirror. We generated 96 sampling positions so that the sample-incident unit vectors were roughly evenly distributed about the unit sphere \cite{deserno2004generate} 
and contained approximately within the nominal angular ranges of $\pm75^\circ$ and $\pm25^\circ$ about the $y$- and $x$-axes, respectively (Fig. 1d,e).
To minimize the probe's total travel distance and time while visiting all 96 positions, we used the 2-opt heuristic to find an approximate solution to the traveling salesman problem, starting at the position closest to that corresponding to normal incidence (Fig. 1d). Further, since the effective focal length (EFL) can vary as a function of incidence angle, the lateral scan range across the sample needed to be dynamically rescaled to obtain approximately the same lateral FOV at the sample. We computed this scale factor numerically under ideal conditions (i.e., perfect paraboloidal shape and alignment) and programmed the galvanometer voltages accordingly. 

Of the 96 multi-view OCT volumes, we used 91 volumes that were successfully registered in the calibration step, described in the next subsection. For fair comparison, we also acquired and averaged 91 repeated OCT volumes from the normal sample-incidence angle, which were first registered to account for any drift.

\subsection{Computational 3D reconstruction}
As described in the main text, the 3D OCRT forward model parameters can be divided into the sample-extrinsic parameters, or the calibration parameters, and the sample-intrinsic parameters, or the sample RI distribution. The calibration parameters control positions and orientations of the final sample-incident rays (400$\times$400$\times$96 total), which we estimated by imaging calibration phantoms consisting of 4-\textmu m polystyrene beads sparsely distributed and embedded in 2\% agarose (w/v). As a first-order estimate, we started with a parametric model, including parameters describing the parabolic mirror focal length, the 6D pose (3D position + 3D orientation) of the 2D plane of probe translation relative to the parabolic mirror, the galvanometer lateral scan amplitudes and 3D orientation (tilt and in-plane rotation), optical dome inner/outer radii and 3D position, and the relative path length difference between the sample and reference arms. All of these parameters were optimizable except for the parabolic mirror focal length and the dome radii, which were left at their nominal values. Since this first-order estimate was insufficient due to misalignment and manufacturing imperfections, we included a nonparametric refinement that allowed all 96 multi-angle OCT volumes to vary individually in terms of their 6D poses, lateral scan amplitudes, and degree of non-telecentricity. In particular, the 3D positions and 2D orientations of the 400$\times$400 rays within each OCT volume were allowed to deviate from the first-order prediction according to six up-to-fourth-order polynomial functions of the nominal 2D lateral coordinates, three for the 3D positions and three for the 2D orientations (normalization of rays to unit vectors reduces dimensionality by one).

These aforementioned calibration parameters define the boundary conditions for ray propagation through the sample, which is parameterized by a 3D RI distribution. Because the sample is roughly index-matched via water immersion, we assume that the rays do not change direction upon propagation, but rather only get delayed. At the cost of bias, this assumption significantly reduces the computational costs of the full solution to the ray equation, as the spatial partial derivatives are no longer needed. Letting $\mathbf{r}_0=(x_0,y_0,z_0)$ and $\mathbf{u}_0=(u_{x,0}, u_{y,0}, u_{z,0})$ be the calibrated starting ray for a given A-scan, the position of the $i^{th}$ pixel of the A-scan, $A_i$, is given by
\begin{equation}\label{rayprop}
    \mathbf{r}_i = \mathbf{r}_{i-1} + \frac{\Delta z}{n(\mathbf{r}_{i-1})} \mathbf{u}_0,
\end{equation}
where $n(\mathbf{r})$ is the 3D RI distribution and $\Delta z$ is the axial pixel sampling period in air. In practice, we use stratified random batching of all the A-scans and their corresponding boundary conditions, whereby the same number of A-scans from each OCT volume are randomly selected, so that we can perform one optimization iteration on a GPU. After propagating all rays in the $j^{th}$ batch, let $\mathbf{A}^\mathit{batch}_j$ be a flattened vector of length $n_A$ containing all the A-scan values from the batch, and $\mathbf{r}^\mathit{batch}_j$ be a flattened matrix of dimensions $n_A\times3$ of the corresponding coordinates after ray propagation. We also optimize a global A-scan background, $\mathbf{A}_\mathit{back}$, a length-$n_A$ vector that is subtracted from every A-scan from every volume, to account for residual background noise stemming, for example, from the OCT source spectrum:
\begin{equation}
    \mathbf{A}^\mathit{batch}_j \leftarrow \mathbf{A}^\mathit{batch}_j - \mathbf{A}_\mathit{back}.
\end{equation}
Then, initializing the reconstruction to a 3D tensor of zeros, $\mathbf{R}_0$, whose size depends on the target 3D FOV reconstruction volume and the voxel size, the weighted moving average estimate of the reconstruction at the $j^{th}$ iteration is given by
\begin{equation}\label{update_rule}
\begin{gathered}
    \mathbf{R}_j\leftarrow\mathbf{R}_{j-1}\\
    \mathbf{R}_j(\mathbf{r}^\mathit{batch}_{j-1}) \leftarrow m\mathbf{R}_{j-1}(\mathbf{r}^\mathit{batch}_{j-1}) +
    (1-m) \mathbf{A}^\mathit{batch}_{j-1},
\end{gathered}
\end{equation}
where $0<m<1$ is a momentum hyperparameter that tunes how quickly to update the moving average reconstruction for each batch; $1-m$ should be on the order of the fraction of all A-scans that are in one batch. Since $\mathbf{R}_j$ is discretized, the nearest $2\times2\times2$ voxel neighborhood surrounding each continuous point in $\mathbf{r}^\mathit{batch}_{j-1}$ is assigned a value according to the trilinear interpolation weights.
This update rule is similar to one we recently proposed for parallax-aware image stitching for photogrammetry \cite{zhou2021mesoscopic}, except the relative batch is much smaller here. As a result, earlier estimates of the reconstruction will be significantly biased towards the zeros initialization, especially since $m$ should be close to 1. This is a similar problem encountered and solved by the Adam optimizer \cite{kingma2014adam}, one of the most commonly used variants of stochastic gradient descent for deep learning applications and the one that we use for our inverse optimization, described below. In our case, instead of correcting the bias in the reconstruction, which would lead to noisy earlier estimates, we correct it in the forward prediction, 
\begin{equation}
    \widetilde{\mathbf{A}}^\mathit{batch}_{j-1} = \frac{\mathbf{R}_j(\mathbf{r}^\mathit{batch}_{j-1})}{1-\widetilde{m}^j},
\end{equation}
where $\widetilde{m}$ is the effective momentum that may be different from $m$, depending on the voxel size of the reconstruction. Given this forward prediction, we quantify how well the current batch of A-scans is registered to the current estimate of the reconstruction via the mean squared error (MSE),
\begin{equation}
    \mathit{MSE}_j=\frac{1}{n_A}||\widetilde{\mathbf{A}}^\mathit{batch}_{j-1}-\mathbf{A}^\mathit{batch}_{j-1}||^2.
\end{equation}
We also included two Tikhonov (L2) regularization terms to aid in the estimation of the 3D RI map, one of which was a standard practice of penalizing the magnitudes of the 3D image gradients to promote smoothness, the other of which was to enforce object support. In particular, for each multi-view OCT volume, we performed a basic intensity-threshold-based segmentation of the first reflection for each A-scan, which corresponds to the tissue surface. During ray propagation (Eq. \ref{rayprop}), the RI values along the ray trajectories are stored if the ray hasn't reached the first tissue boundary. These deviation from these RI values from that of water, the immersion medium, are squared and summed, yielding the object-support-enforcing regularization term. These three terms -- the MSE and two Tikhonov regularization terms -- constitute the loss, which we minimize with respect to all the calibration (sample-extrinsic) and sample-intrinsic (3D RI) parameters via the Adam optimizer \cite{kingma2014adam}.

Optimization was performed using TensorFlow 2.2 on a Google Cloud Platform virtual machine with 6 vCPUs, 32 GB of RAM, and a 16-GB Nvidia Tesla T4 GPU. Each raw 5D OCRT dataset requires about 123 GB of storage, which is streamed to RAM in batches (3,200 batches per epoch).

\subsection{Sample preparation}
All biological samples were fixed in 4\% paraformaldehyde and stored at 4$^\circ$ until ready for imaging. Samples were embedded in 2\% agarose (w/v), mounted upside down at the mirror focus, and immersed in water to avoid having the sample dry out during data acquisition. Animal studies were performed in compliance with Duke University Institutional Animal Care and Use Committee. The PDMS microstamp sample was manually cut into a $\sim$1 mm\textsuperscript{2} and directly immersed in water inverted.

\section{Visualizing the ommatidia of the fruit fly}
Interestingly, the ommatidia are less reflective than structures immediately below it and thus taking a maximum intensity projection (MIP) does not reveal them. Only when we move $\sim$8 \textmu m away from the maximally reflective structures along vectors normal to the fly's surface do the ommatidia become apparent for OCRT, but not OCT (Fig. \ref{fig:fruitfly_eye}c,d).

\section{Generalizing OCRT: alternative multi-angle synthesis strategies in addition to mean reflectivity}
The data collection for 3D OCRT yields a 5D datacube, consisting of OCT backscattered signals as a function of 3D space and 2D orientation, $\mathit{OCT}(x,y,z,k_x,k_y)$. After multi-angle registration, the OCRT reconstructions were formed by averaging the backscattered signals across the angular dimensions, 
\begin{equation}\label{mean}
    \mathit{OCRT}_\mathit{mean}(x,y,z) \propto \iint\limits_{k_x,k_y} \mathit{OCT}(x,y,z,k_x,k_y)\,dk_x\,dk_y,
\end{equation}
which is effective at reducing speckle. However, there are many other possible choices for reducing the dimensionality of the 5D datacube to 3D that would highlight alternate sample features, thereby generalizing OCRT to include new forms of contrast. Thus, Eq. \ref{mean} is a special case of a more general equation,
\begin{equation}
    \mathit{OCRT}_F(x,y,z) = F_{(x,y,z,k_x,k_y)\rightarrow(x,y,z)}\Big\{\mathit{OCT}(x,y,z,k_x,k_y)\Big\},
\end{equation}
where $F$ is an operator that eliminates the angular dimensions, $k_x$ and $k_y$. Whereas $F$ performs the mean operation in Eq. \ref{mean}, other choices of $F$ include variance, higher-order descriptive statistics, and entropy, which would yield more information about the shape of angular backscatter distribution. The argmax function would identify the incidence angle that gives the highest backscattered signal, which would thus produce a 3D orientation map of the sample. 
Taking the Fourier transform across the angular dimensions before their reduction may yield structural information about the sample \cite{wax2002cellular}. 
To identify spatial locations that exhibit a specific angular backscatter distribution profile, another option would be to use template-matching via cross-correlation or mean square error. Data-driven linear or nonlinear dimensionality-reduction strategies could also be leveraged to collapse the angular dimensions, such as principal component analysis (PCA), t-distributed stochastic neighborhood embedding (t-SNE) \cite{van2008visualizing}, or neural networks. Even more generally, in addition to $k_x$ and $k_y$, $F$ could also operate along space and time ($x$, $y$, $z$, $t$). Thus, $F$ would also include spatial convolutions and temporal variance-based approaches, such as OCT angiography (OCTA) \cite{wang2007three} and dynamic OCT \cite{scholler2020dynamic}. 

The wealth of options for mapping the 5D datacube to the 3D OCRT reconstruction opens the door to many future studies that could expand the utility and applicability of 3D OCRT. To demonstrate this potential, we investigated the case where $F$ computes the angular variance,
\begin{equation}\label{variance}
\begin{split}
    \mathit{OCRT}_\mathit{var}(x,y,z) &\propto 
    \iint\limits_{k_x,k_y} \Big(\mathit{OCT}(x,y,z,k_x,k_y) -  \mathit{OCRT}_\mathit{mean}(x,y,z)\Big)^2 \,dk_x\,dk_y,\\
    &= \iint\limits_{k_x,k_y} \mathit{OCT}^2(x,y,z,k_x,k_y)\,dk_x\,dk_y -  \mathit{OCRT}_\mathit{mean}^2(x,y,z).
\end{split}
\end{equation}
where, in practice, we used the second line of Eq. \ref{variance}, as this expanded form only requires two mappings from the OCT datacube space to the OCRT reconstruction space (i.e., $\mathit{OCT}\rightarrow \mathit{OCRT}_\mathit{mean}$ and $\mathit{OCT}^2\rightarrow \mathit{OCRT}^2$), thus allowing us to directly apply our moving average-based reconstruction algorithm described in the Methods section. The first line of Eq. \ref{variance} would have required us to perform an extra inverse mapping step from OCRT space to OCT space (in addition to two forward mappings).
A variance-based OCRT reconstruction would highlight anisotropically backscattering structures, such an oriented flat surface. For example, an ideal spherical particle would yield no variance signal, as the backscattered signal would be independent of incidence angle. However, an ideal mirror would yield a high variance signal, as it strongly retroreflects at normal incidence and doesn't retroreflect at other incidence angles. A good example of orientationally-sensitive structures are those in Henle's fiber layer of the retina \cite{lujan2011revealing}, whose reconstructed OCRT signal could be attenuated if we just use angular averaging (Eq. \ref{mean}).

Fig. \ref{fig:variance} compares 2D cross-sections of the standard-deviation-based 3D OCRT reconstructions of the zebrafish and mouse samples to those of the conventional mean-based 3D OCRT reconstructions. As expected, the standard-deviation-based reconstructions highlight oriented structures, such as the reticular structures within the zebrafish yolk sac, muscle fibers surrounding the mouse esophagus, and the boundaries of the cartilage plates in the mouse trachea, including to some extent the lacuna or the small cavities within the cartilage.

\bibliography{sample}

\begin{figure}
    \centering
    \includegraphics[width=\textwidth]{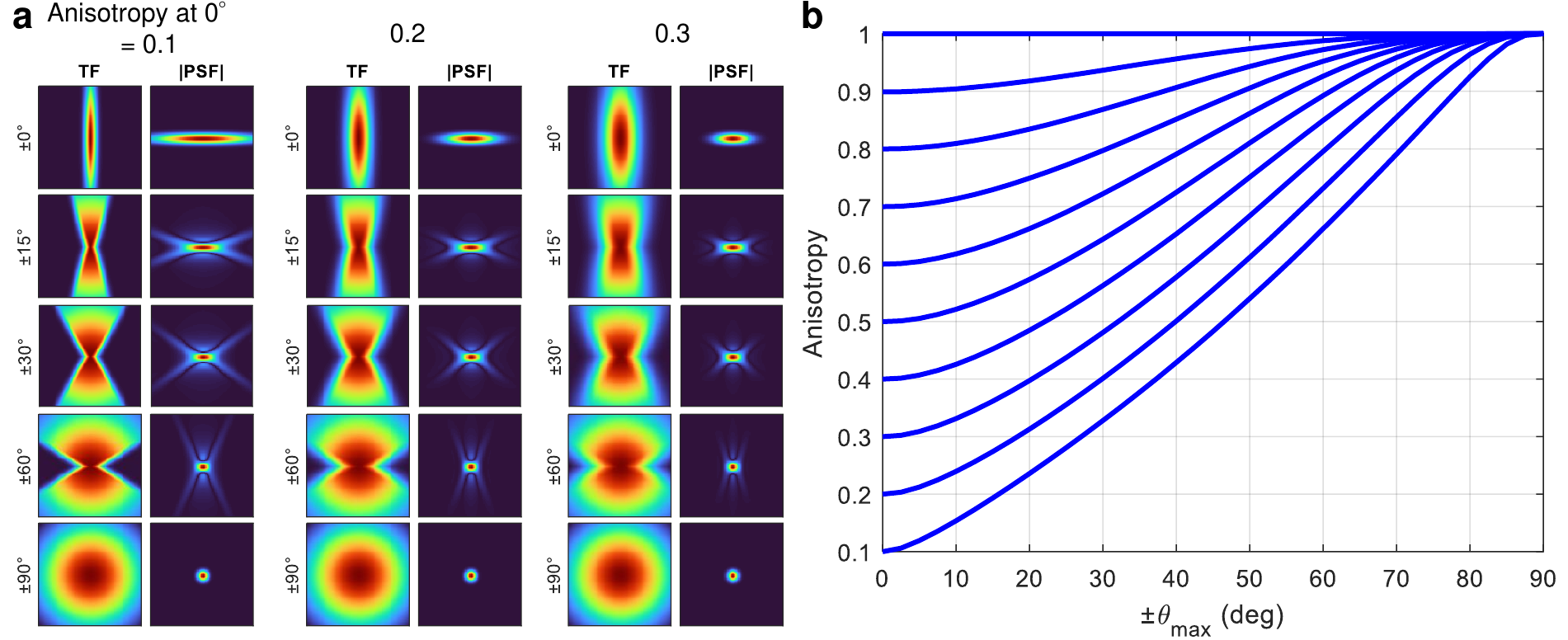}
    \caption{Theoretical resolution enhancement as a function of angular coverage. Resolution is quantified using the full width at half maximum (FWHM) criterion. (a) Simulated transfer functions (TFs) and point-spread functions (PSFs) as a function of angular coverage. The three columns correspond to different resolution anisotropy conditions (axial-to-lateral resolution ratio = 0.1, 0.2, or 0.3). (b) Axial-to-lateral FWHM ratio of synthesized PSF as a function of angular coverage; each curve corresponds to a different initial anisotropy. These curves differ from those of our previous theoretical predictions \cite{zhou2019optical}, which used a different, more conservative resolution criterion.}
    \label{fig:resolution_theory}
\end{figure}

\begin{figure}
    \centering
    \includegraphics[width=.9\textwidth]{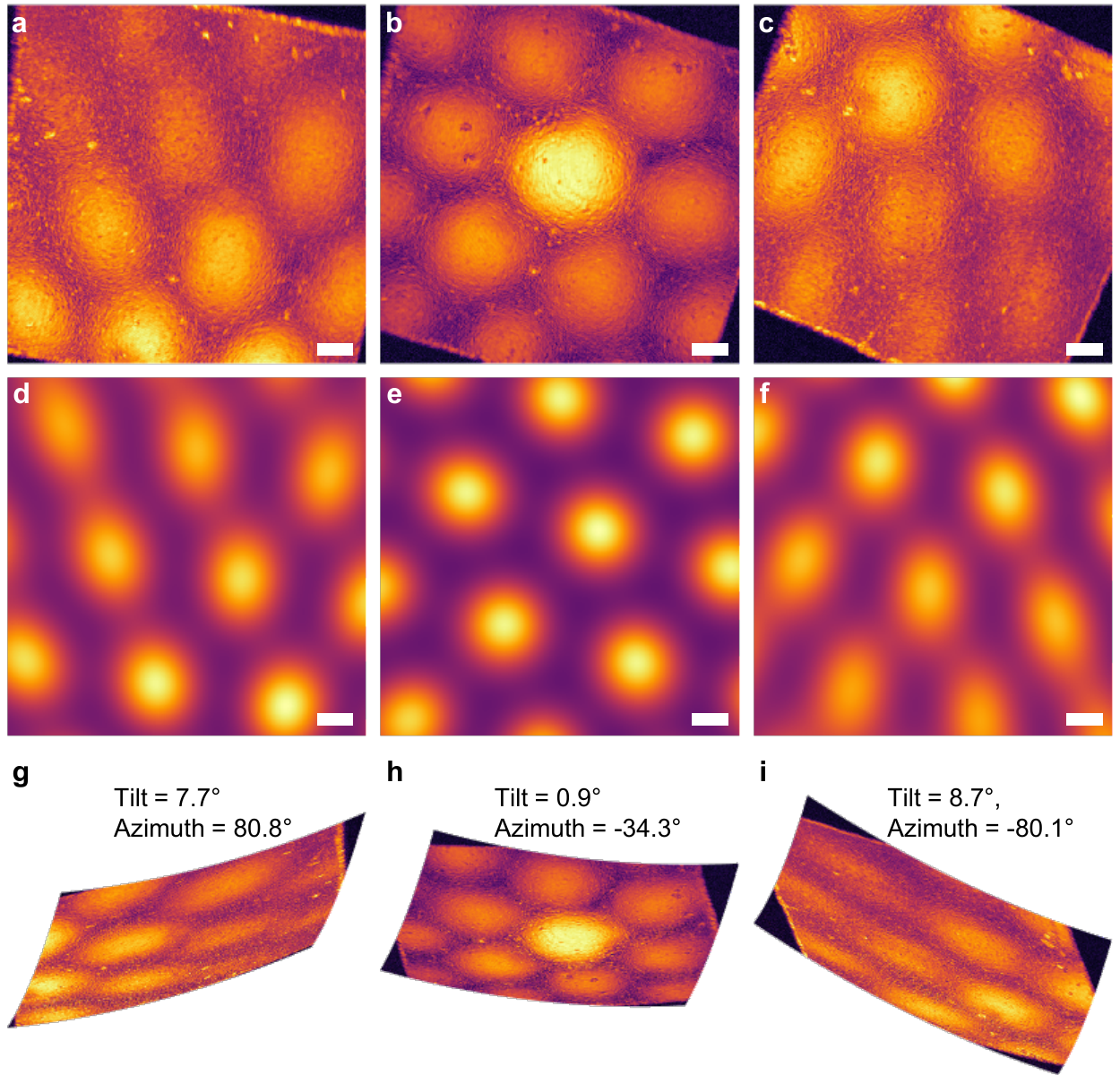}
    \caption{OCT simulations of the interference artifacts of the 5-\textmu m microstamp sample under different tilts match experimental results. (a-c) en face projections of the OCT volumes acquired under three different tilt angles. (d-f) Simulated en face OCT images based on fitting the microstamp surfaces to paraboloids to estimate the local tilts and surface curvature, plotted in (g-i) with a 3$\times$ exaggeration in $z$. The mean orientation of the surface is specified in spherical coordinates, where tilt is with respect to the $z$-axis. Both view angle and non-telecentricity contribute to the interference artifacts. Scale bars, 200 \textmu m.}
    \label{fig:microstamps2}
\end{figure}

\begin{figure}
    \centering
    \includegraphics[width=.8\textwidth]{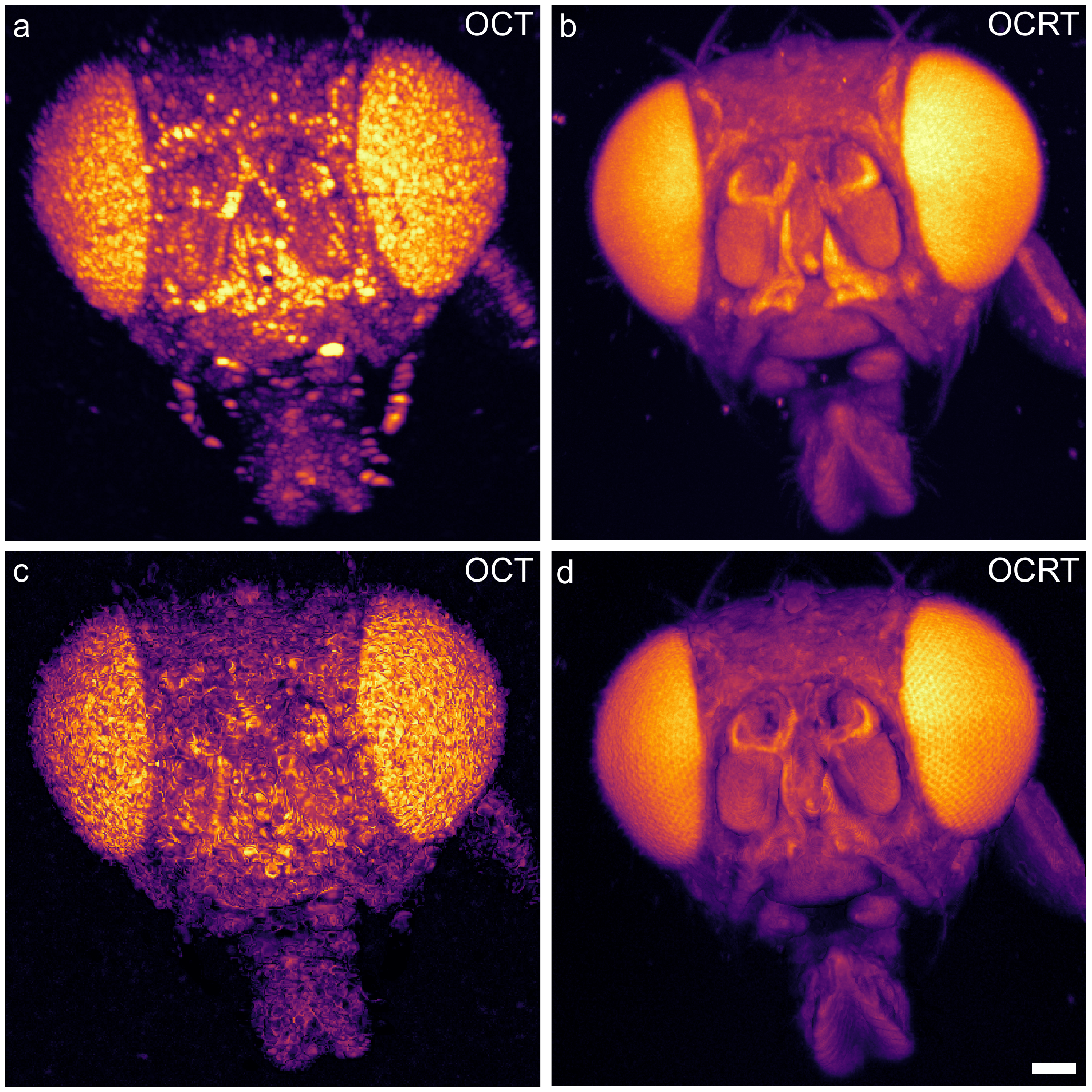}
     \caption{Maximum intensity projection (MIP) comparison of conventional OCT and 3D OCRT reconstruction of a fruit fly head. (a) en face MIP ($z$ points towards the reader) of OCT. (b) en face MIP of the 3D OCRT reconstruction. (c,d) intensity values $\sim$8 \textmu m away, along the surface normals, from the positions corresponding to the maximum intensity for each lateral pixel. At these positions, the micro lenslets (ommatidia) are more obvious in OCRT (zoom into the plot). (a) and (c) are histogram-matched to (b) and (d), respectively. Scale bar, 100 \textmu m.}
    \label{fig:fruitfly_eye}
\end{figure}

\begin{figure}
    \centering
    \centerline{\includegraphics[width=1.3\textwidth]{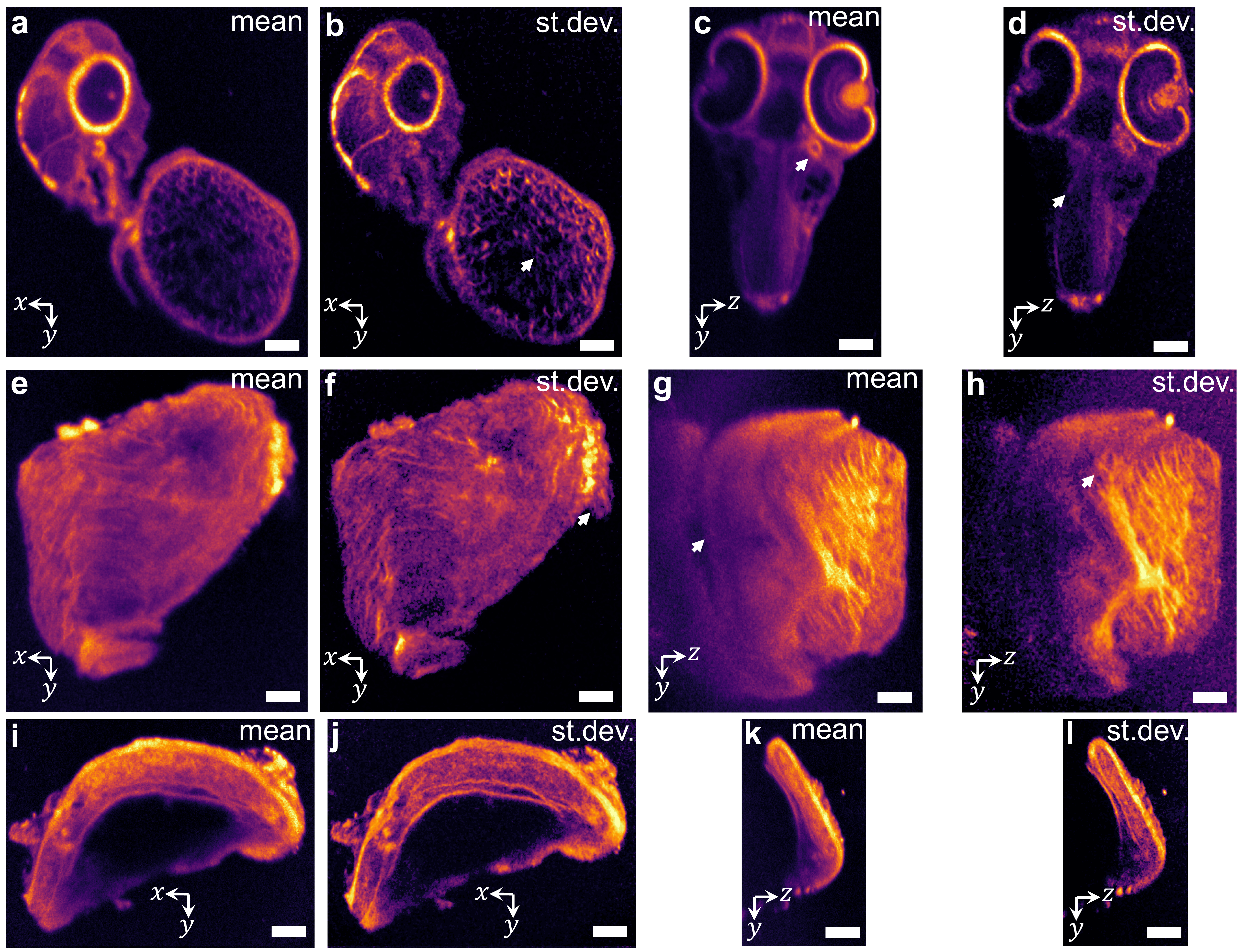}}
     \caption{
     Alternative multi-angle synthesis strategies lead to new contrast mechanisms. The 2D cross-sections of the 3D OCRT reconstructions in (a,c,e,g,i,k) were computed based on the mean OCT reflectivity across all angles, while those in (b,d,f,h,j,l) were computed based on the standard deviation (st. dev.) of the OCT reflectivity across all angles.
     Arrowheads point out a few example features that are highlighted in the mean- or st. dev.-based reconstruction more so than in the other.
     Scale bars, 100 \textmu m.}
    \label{fig:variance}
\end{figure}



